\documentclass[acs,letterpaper, twocolumn]{revtex4}
\usepackage{graphicx}
\usepackage{amsmath}
\usepackage{color}
\usepackage{appendix}

\begin{document}

\title{Super-resolution without Evanescent Waves}

\author{Fu Min Huang and Nikolay I. Zheludev}

\address{Optoelectronics Research Centre, University of Southampton, SO17 1BJ, United Kingdom\\}

\begin{abstract}
The last decade has seen numerous efforts to achieve imaging
resolution beyond that of the Abbe-Rayleigh diffraction limit. The
main direction of research aiming to break this limit seeks to
exploit the evanescent components containing fine detail of the
electromagnetic field distribution at the immediate proximity of
the object. Here we propose a solution that removes the need for
evanescent fields. The object being imaged or stimulated with
sub-wavelength accuracy does not need to be in the immediate
proximity of the superlens or field concentrator: an optical mask
can be designed that creates constructive interference of waves
known as superoscillation, leading to a sub-wavelength focus of
prescribed size and shape in a `field of view' beyond the
evanescent fields, when illuminated by a monochromatic wave.
Moreover, we demonstrate that such a mask may be used not only as
a focusing device, but also as a super-resolution imaging device.
\end{abstract}

\maketitle

The last decade has seen numerous  efforts to achieve imaging
resolution beyond that of the Abbe-Rayleigh diffraction limit,
which proscribes the visualization of features smaller than about
half of the wavelength of light with optical instruments. The main
direction of research aiming to break this limit seeks to exploit
the evanescent components containing fine detail of the
electromagnetic field distribution. Indeed, many powerful concepts
like scanning near-field optical microscopy (SNOM) \cite{1,2}, the
use of various forms of field concentrators \cite{3,4,5,6} and
superlenses \cite{7,8,9,10,11,12,13} depend on the exploitation or
recovery of evanescent fields. The proper far-field optical
superlens \cite{7} requires bulk negative index materials that are
still to be developed, while other designs, though offering
substantial advances, are united by a common severe limitation:
that the object being imaged or stimulated must be in the
immediate proximity of the superlens or field concentrator. Here
we propose a solution that removes this limitation: an optical
mask can be designed that creates constructive interference of
waves known as superoscillation, leading to a sub-wavelength focus
of prescribed size and shape in a `field of view' beyond the
evanescent fields, when illuminated by a monochromatic wave.
Moreover, we demonstrate that such a mask may be used not only as
a focusing device, but also as a super-resolution imaging device.

In fact, in his seminal  1952 paper Toraldo Di Francia
demonstrated that propagating waves can create sub-wavelength
localization of light in the far-field with the suggestion of a
pupil design providing an accurately tailored sub-wavelength
diffraction spot using a series of concentric apertures \cite{14}.
Such a sub-wavelength concentrator could be employed as a focusing
device in a super-resolution scanning optical microscope where the
object is placed several wavelengths away from the device, thus
removing the main limitation of near-field instruments. More
recently, Berry and Popescu \cite{15}, starting from earlier works
on quantum mechanics, showed that diffraction on a grating
structure could create sub-wavelength localizations of light that
propagate further into the far field than more familiar evanescent
waves. They relate this effect to the fact that band-limited
functions are able to oscillate arbitrarily faster than the
highest Fourier components they contain, a phenomenon now known as
superoscillation \cite{16,17,18}. Examples of sub-wavelength
localizations of light generated by a nano-hole array \cite {19,20} and a
thin meta-dielectric shell \cite{21} have been demonstrated recently.  Research on beating the diffraction limit
actually has an even longer history: in 1922, Oseen, with
reference to Einstein's radiation `needle stick', proved that a
substantial fraction of the emitted electromagnetic energy can be
sent into an arbitrarily small solid angle \cite{22}. Beginning
from the pioneering work of Shelkunoff \cite{23}, the microwave
community contemplated the idea of achieving antennae that beat
the diffraction limit for directivity: several authors were able
to prove that for a linear array of properly adjusted radiating
antenna dipoles, there were no theoretical limits to directivity
whatsoever \cite{24,25}. However, the sharp increase in the
proportion of reactive to radiated power that would be required to
achieve super-directivity means that the antenna gain increase is
offset by the need to provide an even higher increase in the power
to the antenna to maintain the signal level, thus rendering the
concept of super-directive antennae impractical. As we will see
below, achieving a sub-wavelength localization of light in the
far-field also comes at a price of losing most of the optical
energy into diffuse sidebands. Nevertheless, optical microscopy
applications can tolerate much higher losses than those acceptable
in antenna design: scanning microscopes can work with only a few
photons per second, giving one around 19 orders of magnitude of
power reserve (assuming that a 1 W laser is used as the optical
source).

\begin{figure}[h]
\includegraphics[width=80mm]{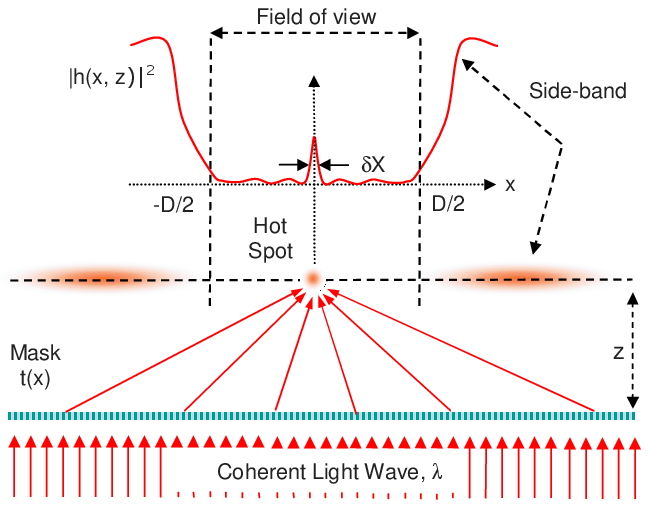}
\caption{Optical superoscillation. An arbitrary field distribution with an arbitrarily small hot-spot $\delta x$, within a limited area $[-D/2,D/2]$, can be generated by diffraction of a plane electromagnetic wave from a purposely designed mask.}
\label{configuration}
\end{figure}

In this letter we make a further step in the investigation of the potential of superoscillation for imaging and achieving sub-wavelength foci. We derive an algorithm for designing a mask that creates a sub-wavelength focus of prescribed size and shape within a prescribed `field of view' when illuminated by a monochromatic wave. Moreover, we show that such a mask may be used not only as a focusing device, but also as a super-resolution imaging tool. We also study the role of manufacturing imperfections on the achievable super-resolution and suggest a design for a superoscillation plasmonic energy concentrator.

The typical situation that we address here is presented in Fig.\ref{configuration}: we aim to design a mask which, within a limited area $[-D/2,D/2]$ (field of view), will create a small hot-spot of light concentration (superoscillation) with a width $\delta x$ located outside the evanescent zone, at a distance $z>\lambda$ from the mask, where $\lambda$ is the wavelength of light illuminating the mask. We argue that in principle a mask can be designed to create a hot-spot that is arbitrarily small, with an arbitrary profile, located at any given distance from the mask.

\begin{figure}[h]
\includegraphics[width=80mm]{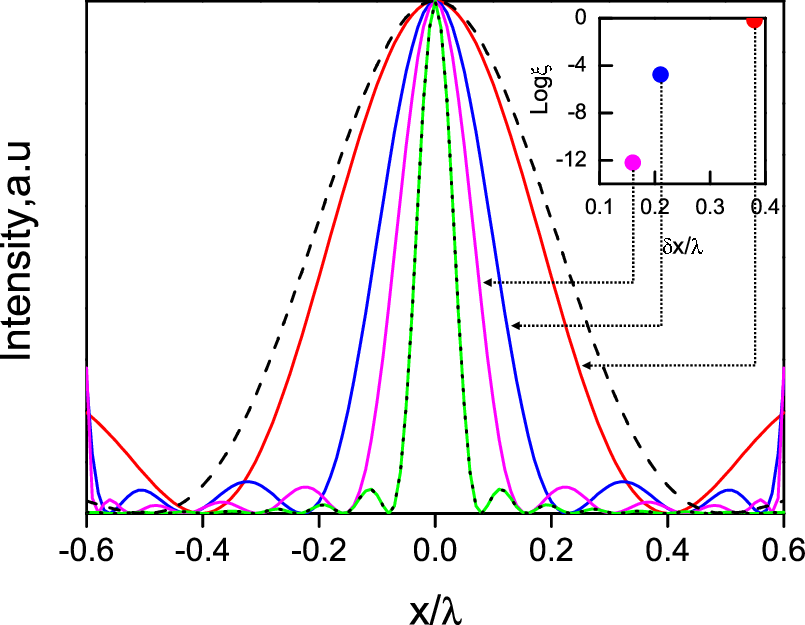}
\caption{Normalized intensity profiles for a hot-spot with $\delta x$ = $0.07 \lambda$ (black dotted line) and its approximations by series of bandwidth-limited prolate spheroidal wave functions $|h(x)|^2$ truncated at: $N=2$ (red line, $\delta x_2$=$0.38 \lambda$); $N=6$ (blue line, $\delta x_6$=$0.21 \lambda$); $N=10$ (pink line, $\delta x_{10}$=$0.14 \lambda$); $N=26$ (green line, $\delta x_{26}$ = $0.07 \lambda$ ). The dashed line shows the $\lambda/2$ limit for a hot-spot achievable with a high-numerical aperture cylindrical lens. Insert: Energy ($\xi$) contained within the hot-spot as a function of its size ($\delta x$)}
\label{super}
\end{figure}

Our consideration is limited to a one-dimensional mask $t(x)$ creating a one-dimensional sub-wavelength field distribution $h(x)$ (superoscillation) when illuminated by plane wave at wavelength $\lambda$, although generalization to a two-dimensional case is trivial. The desired superoscillation feature located at a distance from the near-field zone of the mask can only be created by diffraction on the mask if the feature can be decomposed into a series of plane waves with wave-vector $|\textbf{k}_{0}|=\frac{2\pi}{\lambda}$. Therefore, the main step in designing the mask is to present the desired superoscillating field as a series of bandwidth limited functions that can be decomposed into free-space plane waves of the given wavelength $\lambda$. We argue that \emph{any} arbitrary small field feature can be presented as a series of \emph{band-limited} functions if we are concerned with a prescribed `field of view' $[-D/2, D/2]$. This may be achieved using the formulism of prolate spheroidal wave functions developed by Slepian and Pollark \cite{26} to treat problems of information compression. This is a complete set of functions orthogonal in the interval $[-D/2, D/2]$ and across the whole range $[-\infty, \infty]$. The main feature of prolate spheroidal wave functions is that they are \emph{band-limited} to a frequency domain $[-k_{0},k_{0}]$. Therefore the mask design algorithm comprises the following steps: initially the desired sub-wavelength hot-spot is presented as a series of prolate spheroidal wave functions, which can be truncated when a satisfactory level of approximation is achieved; at the second step this series of prolate spheroidal wave functions is presented as a series of plane waves and, using the scalar angular spectrum description of light propagating from the mask to the super-oscillating feature, the required complex mask transmission function $t(x)$ can be readily derived. The formulism of the algorithm is presented in the Appendix.

\begin{figure}[h]
\includegraphics[width=80mm]{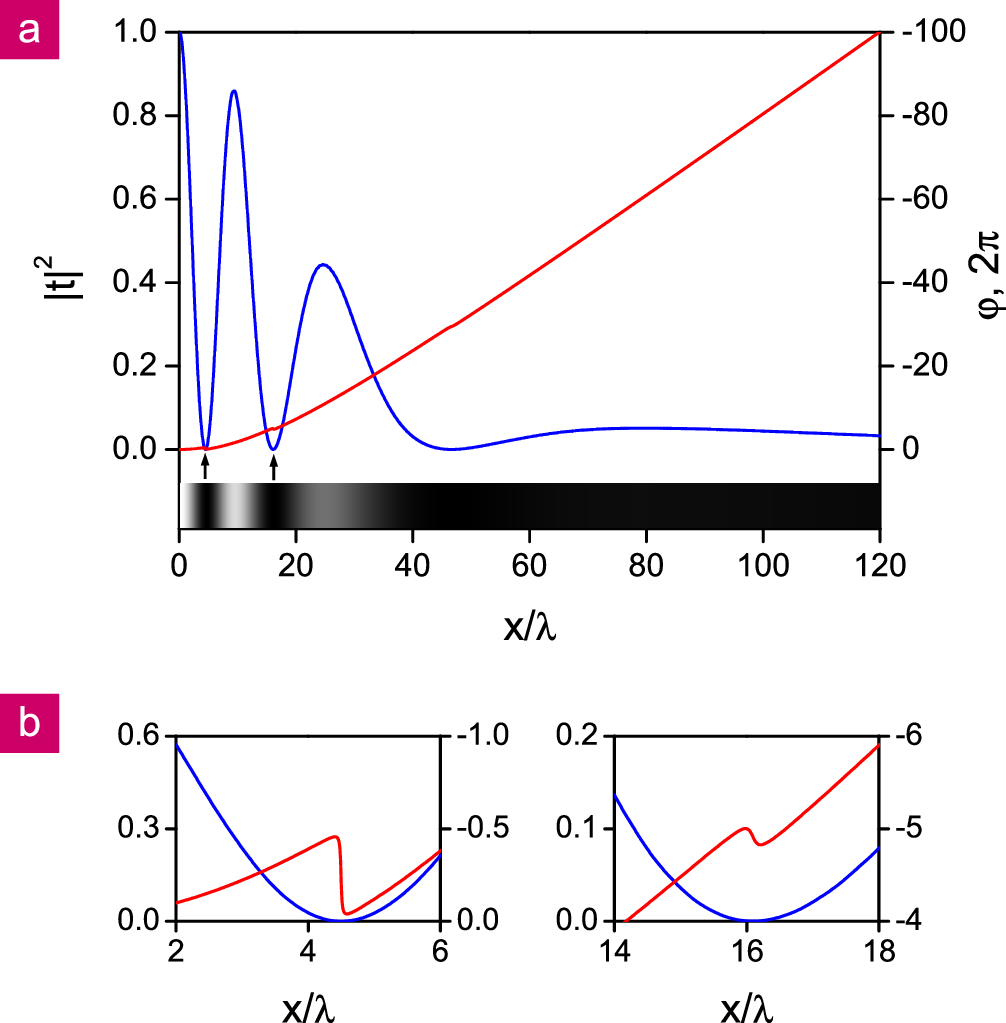}
\caption{Mask profile for the generation of a sub-wavelength hot-spot. (a) shows intensity $|t(x)|^2$ (blue line) and phase $\varphi(x)$ (red lines) profiles of the mask transmission function $t(x)$, which generates a hot-spot with $\delta x$ = $0.21 \lambda$ at a distance $z= 20 \lambda$ from the grating. At the bottom, a grey-scale map of the intensity profile is also presented. (b) shows close-up detail of kinks in the phase curve at the points highlighted with arrows in (a).}
\label{grating}
\end{figure}

In what follows, we will give an explicit example of a mask designed to generate a sub-wavelength concentration of light. Let us aim for a single hot-spot field distribution $Sinc(ax/\lambda)$ centered in the `field of view' $[-D/2,D/2]$. The full width at half maximum $\delta x$ of the intensity profile of this distribution is measured as $2.784\lambda/a$. From now on we will set $a=40$, aiming therefore to achieve a $\delta x = 0.07 \lambda$ hot spot in a $D=1.2 \lambda$ field of view, far beyond the Abbe-Rayleigh limit. Fig.~2 shows the intensity profile of this distribution alongside a number of consecutive approximations to the distribution formed by limited series of prolate spheroidal wave functions, and a curve representing the Abbe-Rayleigh limit that would be achievable by a high-numerical aperture cylindrical lens. One can see that the series rapidly converges and that when $N=26$ the width of the approximation is practically the same as that of the target field distribution.

Fig.3 shows the intensity $|t(x)|^2$ and phase $\varphi$ profiles of the mask  $t(x) = |t(x)|e^{i\varphi}$ required to create the field profile corresponding to $N=6$, which has $\delta x$ = $0.21 \lambda$, at a distance $z= 20 \lambda$ from the grating. As the mask transmission function is even, only part of profile for $x \geq 0$ is shown here. One can see that there is a small central area $[-40 \lambda, 40 \lambda]$ of the mask that transmits most of light. Beyond that area, there is a low-intensity broad shoulder of a slowly fading transmission characteristic. The phase profile of the mask resembles that of a concave lens where the optical thickness is at a minimum in the centre. The monotonous increase is only interrupted at a few positions where transmission amplitude is close to zero (indicated by arrows) and the phase shows a kink as illustrated in the zoomed sections $(b)$.

\begin{figure}[h]
\includegraphics[width=80mm]{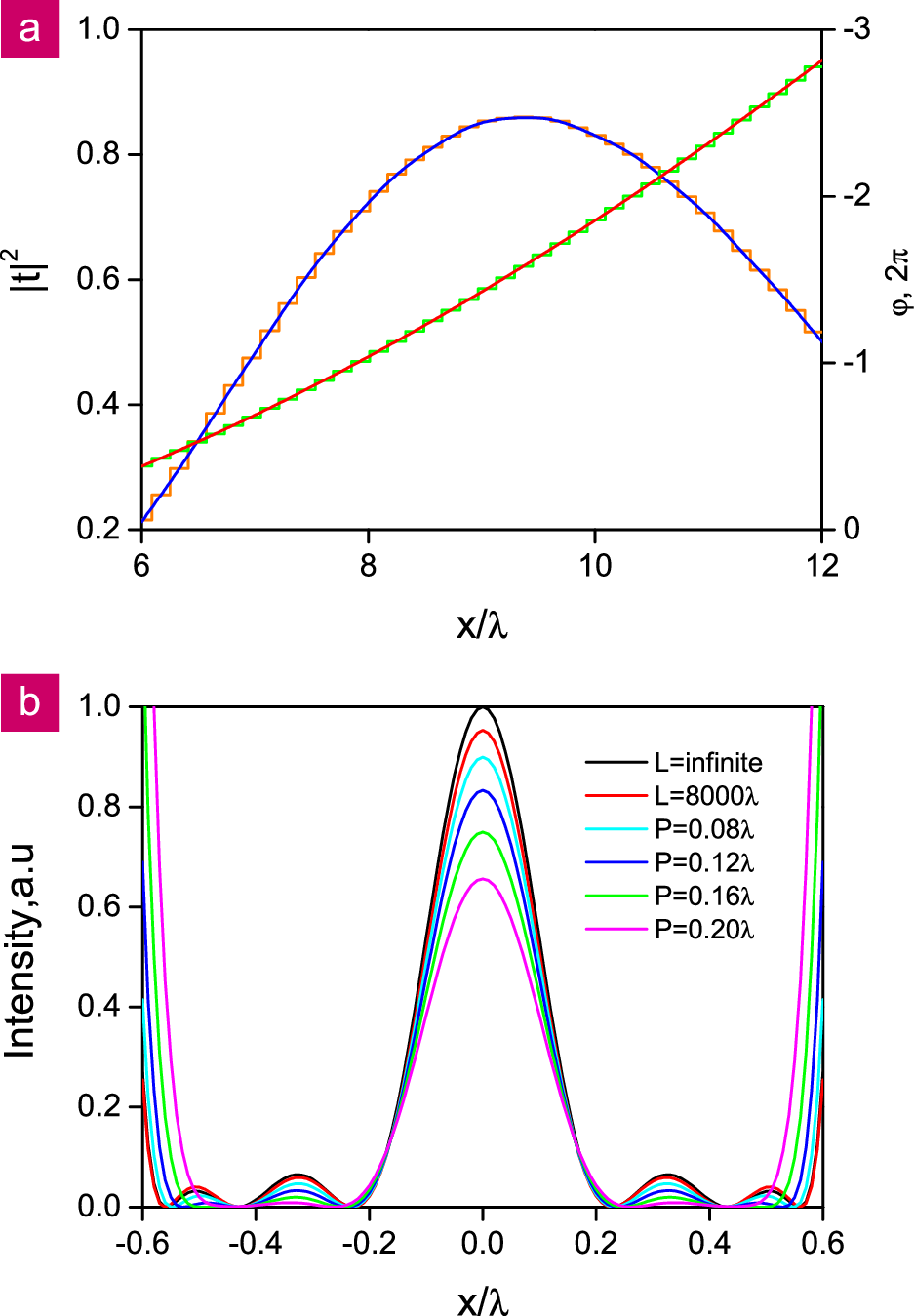}
\caption{The role of finite mask length and manufacturing imperfections on mask performance. (a) shows the intensity and phase characteristics of an ideal mask (smooth line) and its stepped equivalent, where the step of $0.16 \lambda$ mimics the manufacturing pixelation. (b) Calculated intensity profiles of hot-spots generated by an ideal infinitely long mask (black line); an ideal mask truncated to a length of $8000 \lambda$ (red line); and $8000 \lambda$ long masks pixelated to $0.08 \lambda$ (cyan line), $0.12 \lambda$ (blue line), $0.16 \lambda$ (green line ) and $0.2 \lambda$ (pink line).}
\label{field}
\end{figure}

\begin{figure}[h]
\includegraphics[width=80mm]{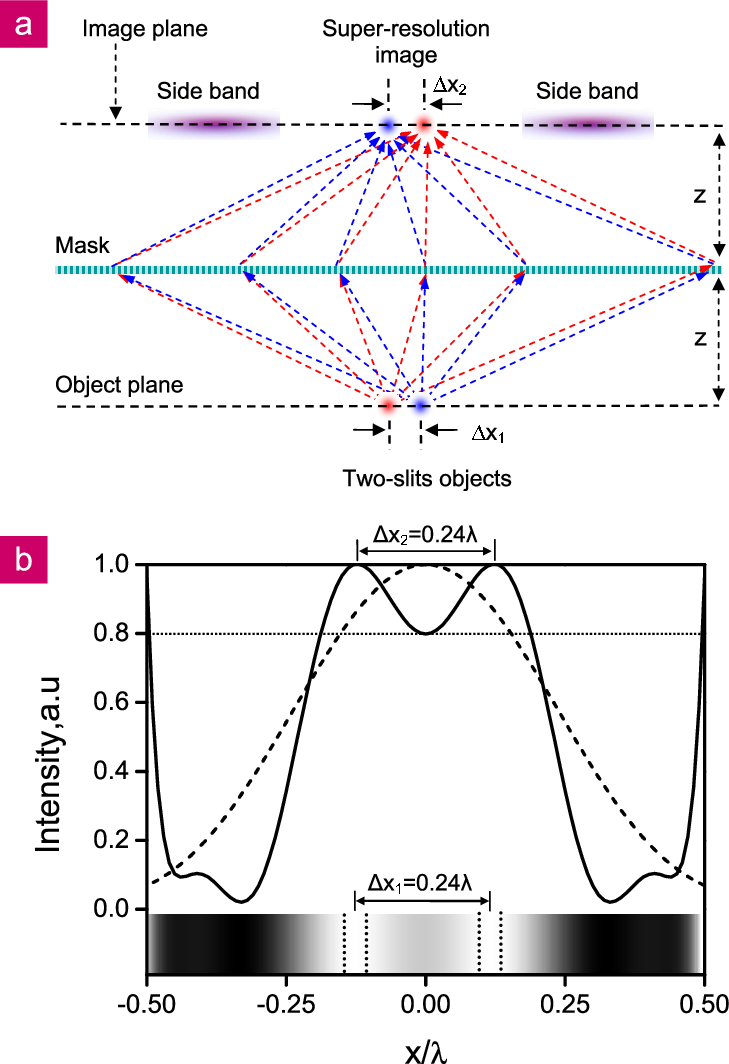}
\caption{(a) Superoscillation mask as a sub-wavelength imaging device. (b) Intensity profile of the image of two sub-wavelength slits separated by a distance $\Delta x_1 = 0.24 \lambda$, located at a distance $20 \lambda$ from the mask. The intensity distribution is also shown in gray-scale at the bottom of the image where the dotted lines indicate the positions of the two slits. For comparison, the dashed line shows the intensity profile of the image field of the two slits as seen imaged by a conventional cylindrical lens with unitary numerical aperture and a resolution of $\lambda/2$.}
\label{imaging}
\end{figure}

The superoscillation process is based on the precise and delicate
interference of waves, so we investigated its stability with
respect to the manufacturing tolerances to which masks can be
fabricated and to the need to use masks of finite length (see the
description of the practical device design below). We found that
for the mask presented in Fig.~3 an $8000 \lambda$ long device may
be used instead of an infinitely long mask without any substantial
degradation of performance ($\delta x$= $0.21 \lambda$ for both
infinite and truncated masks with a 5\% intensity decrease for the
finite mask). A practical mask is likely to be manufactured by
electron-beam lithography with a certain limited resolution, so we
also investigated the dependence of superoscillation hot-spot size
and intensity on the pixelation of the grating design: we replaced
the smooth transmission function with a stepped functions in which
the step width corresponded to the resolution $P$ of the
manufacturing process (see Fig.~4a). Fig.~4b shows the variation
of the intensity profile with increasing pixelation. The
superoscillation process is remarkably stable against this
manufacturing imperfection: with pixel size increasing to $0.2
\lambda$, the $\delta x$ width of the hot-spot only increases by
about 8\%. Pixelation has a more substantial influence on the peak
intensity, which is reduced by about 40\% under such conditions.

To quantitatively characterize the optical energy contained in a
superoscillating hot-spot, we define $\xi$ as the ratio between
the energy contained in the hot-spot (between the first intensity
minima either side of the central peak) and the total energy
transmitted through the mask. In the inset to Fig.\ref{super} we
show values of $\xi$ against the width of the peak $\delta x$ for
different levels of approximation $N=2,6,10$. One can clearly see
that the smaller the hot-spot is, the lower the proportion of
energy that goes into it. For example, for $N=6$ only
$1.74\times10^{-5}$ of the total transmitted energy will be
focused into the hot-spot. The energy contained in the hot-spot is
also significantly dependent on the field of view: if the required
field of view is decreased by a factor of two from $D=1.2 \lambda$
to $D=0.6 \lambda$, the proportion of energy going into the
hot-spot at $N=6$ increases by nearly three orders of magnitude to
$4.8 \times 10^{-2}$. In fact, a more general consideration of the
super-oscillating functions shows that the intensity in the
hot-spot may only decrease polynomially with its width \cite{17}.
We therefore argue that when designing practical realizations of
super-oscillating masks, less demanding requirements on the shape
of hot-spots may yield much higher values of $\xi$. In particular,
one may achieve much higher values of $\xi$ for a given size of
hot-spot by relaxing the requirements on the field structure of
the hot-spot `pedestal' (the background level within the field of
view), for instance by stipulating only the maximum field
intensity in the pedestal area relative to that of the hot-spot,
but not the exact profile of the pedestal.

Recently it was shown that a hole array can be used as an imaging
device \cite{27}. Similarly, a super-oscillating mask can also be
designed to image a sub-wavelength object. This is illustrated in
Fig.~5 where we show a mask with a transmission function that
converts an object located $20 \lambda$ from the mask to an image
on the opposite side of the mask, also at $20 \lambda$ (see
formula $A6$ of the Appendix). Fig. ~5(b) illustrates the imaging
of two incoherent slit sources (each $0.04 \lambda$ wide)
separated by a distance of $0.24 \lambda$. Here the slits are
clearly resolved according to the Rayleigh criterion \cite{28},
which state that the total intensity at the saddle point of the
sum intensity profile of two just-resolved slit sources is 81\% of
the maximum intensity.

Regarding practical  implementations of the superoscillating mask,
the above calculations demonstrate that manufacturing tolerance
should be of the order of $0.1 \lambda$. Thus, manufacturing such
a phase mask from a slab of dielectric material for microwave and
THz frequency focusing should not be a very challenging problem.
It can then be covered with an absorbing film of variable
thickness to create the desired transmission profile. Fabricating
such a mask for the IR and optical parts of the spectrum presents
a more significant challenge and a fabrication accuracy of between
$5$ and $50$nm will be required. However, even such challenging
phase masks may be created from glass with the diamond milling
techniques used to produce aspheric lenses, while further
fine-tuning with true nanoscale resolution may be achieved using
focusing ion beam milling. Here the absorbing part of the mask
could be a metal film of variable density prepared by UV or e-beam
lithography.

\begin{figure}[h]
\includegraphics[width=80mm]{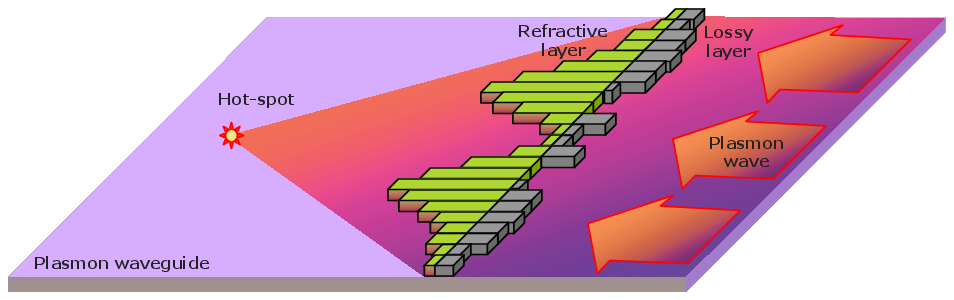}
\caption{Plasmonic superoscillation focusing. A plasmonic mask is archived by placing a sculptured dielectric refractive layer on top of a flat metal surface to create the phase profile followed by a sculptured lossy layer to create the intensity profile.}
\label{plasmon}
\end{figure}

The idea of  superoscillation could also be applied to the
creation of a super-high-resolution plasmonic device. Surface
plasmon-polaritons are collective oscillations of light and
electrons that propagate along the interface between a metal (e.g.
gold, silver) and a dielectric. They are essentially
two-dimensional waves that can create complex field patterns by
interference \cite{29,30}. As plasmon wavelengths are somewhat
shorter than those of light at the same frequency, plasmonic
devices promise better resolution than optical devices even within
the conventional Abbe-Rayleigh diffraction limit. We argue that
this may be enhanced even further through the use of
superoscillation as the approach developed above may be easily
applied to plasmons. A possible implementation of a plasmonic
focusing device is presented in Fig.~6. Here, in order to create a
desired intensity and phase profile of the mask we exploit the
fact that the complex refractive index for plasmons depends on the
dielectric properties $\varepsilon$ of the medium forming an
interface with the metal \cite{31}:
$\tilde{n}=\sqrt{\frac{\varepsilon_{m}\varepsilon}{\varepsilon_{m}+\varepsilon}}$.
By preparing a film of lossless dielectric on a metal surface with
a prescribed profile, one can create the desired phase profile for
the plasmon mask. Using an additional lossy profiled dielectric
layer on the top of the first the necessary intensity profile can
be engineered. Here the transmission function of the surface
plasmon mask should be derived taking account of losses (see
Appendix, formula A7).

In summary,  we have shown that an optical mask can be designed
that creates a a sub-wavelength focus in an area beyond the
evanescent fields. Such a mask may also be used  as a
super-resolution imaging device with numerous applications for
instance for imaging inside a leaving cell which is impossible
with a near-field device.

\begin{acknowledgments}
The authors would like to acknowledge the financial support of the Engineering and Physical Sciences Research Council, UK.
\end{acknowledgments}

.

\appendix
\section{Designing a superoscillating mask}

Here we describe a  method to design a mask with a complex
transmission function $t(x)$ that will generate a prescribed field
distribution $f(x)$ within a limited region $[-D/2, D/2]$ at a
distance $z$ from the mask using prolate spheroidal wave functions
\cite{26}. We assume that the mask is illuminated at normal
incidence with a plane monochromatic wave $E(x,z=0) = 1$ (the time
dependent factor $e^{i\omega t}$ is omitted) with a wavelength
$\lambda=2\pi/k_{0}$. In the scalar angular spectrum description
of light propagation \cite{32} the field at a point $(x,z)$ is

\begin{equation}
E(x,z)=
\int^{k_{0}}_{-k_{0}}T(u)e^{iux}e^{iz\sqrt{k^{2}_{0}-u^{2}}}du
\label{exz}
\end{equation}

where $T(u)$ is the Fourier transform of $t(x)$. We now approximate $h(x)\equiv E(x,z)$ as a limited series of orthogonal prolate spheroidal wave functions $\psi_{n}(c,x)$ that are band-limited to the frequency domain $[-k_{0},k_{0}]$

\begin{equation}
h_N(x)=\sum_{n=0}^{n= N}{a_{n}(c)\psi_{n}(c,x)} \label{hx}
\end{equation}

Here $a_{n}$ depends on a constant $c=\frac{\pi D}{\lambda}$ and the Fourier transform function of $h_{N}(x)$ is given by

\begin{equation}
H_{N}(u)=\sum_{n=0}^{n=N}{\frac{\pi a_{n}\psi_{n}(c,\frac{u D}{2k_{0}})}{i^{n}R_{0n}^{(1)}(c,1)}}
\label{Fourier2}
\end{equation}

where $R_{0n}^{(1)}(c,1)$ is a radial prolate spheroidal wave function of the first kind and

\begin{equation}
h_{N}(x)=\int^{k_{0}}_{-k_{0}}H_{N}(u)e^{iu x}du
\label{hN2}
\end{equation}

Comparing formulae (\ref{exz}) and (\ref{hN2}), we find that the required transmission function $t(x)$ of the mask is

\begin{equation}
t(x)= \sum_{n=0}^{n=N}{\int^{k_{0}}_{-k_{0}} \frac{\pi
a_{n}\psi_{n}(c,\frac{u
D}{2k_{0}})}{i^{n}R_{0n}^{(1)}(c,1)}e^{iux-iz\sqrt{k_{0}^{2}-u^{2}}}}du
\label{Fourier}
\end{equation}

It follows from the angular spectrum representation that to design a mask with a transmission function $m(x)$, which upon illumination by a single slit source (located at a distance $z_{1}$ from the mask), will convert its divergent incident field $E_{0}(x)$ into a prescribed field distribution $h(x)$ at a distance $z$ from the other side of the mask, the following formula should be used:

\begin{eqnarray}
m(x)=\frac{t(x)}{\int^{k_{0}}_{-k_{0}} F(u)e^{iux}e^{iz_{1}\sqrt{k^{2}_{0}-u^{2}}}du} \\
\nonumber
\label{E0x}
\end{eqnarray}

where $F(u)$ is the Fourier transform function of the slit source.

In the case of lossy medium the transmission function $A(5)$ shall
be corrected:

\begin{eqnarray}
\nonumber t(x) &=& \sum_{n=0}^{n=N}{\int^{|k_{0}|}_{-|k_{0}|}\frac{\pi a_{n}\psi_{n}(c,\frac{u D}{2|k_{0}|})}{i^{n}R_{0n}^{(1)}(c,1)}}e^{z Im[\sqrt{k^{2}_{0}-u^{2}}]} \\
&\times& e^{iux-iz Re[\sqrt{k^{2}_{0}-u^{2}}]}du
\label{plasmonfunction}
\end{eqnarray}

where $k_{0}$ is a complex wave-vector in the lossy medium.

\end{document}